\documentclass{svjour3}     
\smartqed 
\usepackage{graphicx}
\usepackage{float}
\usepackage{multirow}
\usepackage{amsmath}
\usepackage{mathptmx}
\usepackage{latexsym}

\journalname{Nonlinear Dynamics}
\begin{document}
\title{Multiple resonance and anti-resonance in coupled Duffing oscillators}
\titlerunning{Multiple resonance and anti-resonance in coupled Duffing oscillators} 
\author{R. Jothimurugan \and K. Thamilmaran \\ S. Rajasekar \and M. A. F. Sanju{\'a}n}
\institute{R. Jothimurugan \and K. Thamilmaran$^*$\at
	      Centre for Nonlinear Dynamics\\
              School of Physics\\
	      Bharathidasan University\\
	      Tiruchirappalli - 620 024\\ Tamilnadu, India \\
              \email{jothi@cnld.bdu.ac.in}\\
	      \email{$^*$maran.cnld@gmail.com}
	\and
	   S. Rajasekar \at
	      School of Physics\\
	      Bharathidasan University\\
	      Tiruchirappalli - 620 024\\ Tamilnadu, India \\
	      rajasekar@cnld.bdu.ac.in   
        \and
           M. A. F. Sanju{\'a}n \at
	     Nonlinear Dynamics, Chaos\\
             and Complex Systems Group\\
	     Departamento de F{\'i}sica\\
             Universidad Rey Juan Carlos\\
 	     Tuli{\'p}an s/n, 28933 M{\'o}stoles\\
             Madrid, Spain\\
             \email{miguel.sanjuan@urjc.es}      
}
\date{Received: date / Accepted: date}
\maketitle
%
%
\begin{abstract}
We investigate the resonance behaviour in a system composed by $n$-coupled Duffing oscillators where only the first oscillator is 
driven by a periodic force, assuming a nearest neighbour coupling. We have derived the frequency-response equations for a system 
composed of two-coupled oscillators by using a theoretical approach. Interestingly, the frequency-response curve displays two 
resonance peaks and one anti-resonance. A theoretical prediction of the response amplitudes of two
oscillators closely match with the numerically computed amplitudes. We analyse the effect of the coupling strength on the 
resonance and anti-resonance frequencies and the response amplitudes at these frequencies. For the $n$-coupled oscillators 
system, in general, there are $n$-resonant peaks and ($n-1$) anti-resonant peaks. For large values of $n$, except for the 
first resonance, other resonant peaks are weak due to linear damping. The resonance behaviours observed in the $n$-coupled 
Duffing oscillators are also realized in an electronic analog circuit simulation of the equations. Understanding the role of coupling 
and system size has the potential applications in music, structural engineering, power systems, biological networks, 
electrical and electronic systems.
%
%
\end{abstract}
%
\section{\label{s1}Introduction}
The typical frequency-response curve of a linear or nonlinear oscillator with a single degree of freedom subjected to an additive 
periodic driving force with a single frequency displays a single resonance peak. Furthermore, when the system is linear and undamped, 
the response amplitude becomes a maximum when the frequency of the driving force matches with the natural frequency of the system. 
In other oscillators a single resonance peak occurs at a frequency different from their natural frequencies. For a system of 
$n$-coupled linear oscillators where only the first oscillator is driven by an additive periodic force, for certain types of 
coupling, the frequency-response curve of each oscillator exhibits at most $n$ peaks depending upon the values of the parameters of the oscillators \cite{r1}. The peaks are the resonances and the 
corresponding frequencies are the resonant frequencies. The valleys in the frequency-response curve are
the anti-resonant frequencies. There are $n-1$ anti-resonant frequencies. In the absence of damping and for driving frequencies 
equal to the anti-resonant frequencies, the response amplitude vanishes. The multiple resonance and anti-resonance phenomena occur 
in nonlinear systems also as shown in the present work.

\indent In the recent years resonance is investigated on coupled systems also. For example, the coupling can enhance 
coherence resonance (CR) in Hodkin-Huxley neuronal network \cite{r13a}. 
The onset and control of stochastic resonance in two mutually  coupled driven bistable systems subjected to independent noises 
are investigated \cite{r13c}.  In a coupled bistable system, coupling improves the reliability of the logic system and thus enhances 
the logical stochastic resonance effect. Moreover, enhancement is larger for larger system size, whereas for large enough 
size the enhancement seems to be saturated \cite{r13d}. System size resonance and coherence resonance are 
demonstrated using coupled noisy systems, Ising model \cite{r13e} and FitzHugh-Nagumo model \cite{r13f}. 
Effect of vibrational resonance of neuronal systems depends extensively on the network structure and parameters, such
as the coupling strength between neurons, network size, and rewiring probability of single small world networks, 
as well as the number of links between different subnetworks and the number of subnetworks in the modular networks \cite{r13g}. 
The enhanced signal propagation is achieved in the coupled Duffing oscillators in the realm of ghost vibrational resonance \cite{r13h}. 
Mechanisms of fano resonances is demonstrated by both theoretically and experimentally in coupled plasmonic system \cite{r13i}. 
Very recently, the influence of nonlinearities on the collective dynamics of coupled Duffing-van der Pol oscillators subjected to
both parametirc and external perturbations has been reported \cite{r13j}.

\indent Why is the study of anti-resonance important? What are its real practical applications? Details of anti-resonance are 
useful in the design of chemotherapeutic protocols \cite{r2}, dynamic model updating \cite{r3,r4,r5,r6} and desynchronizing 
undesired oscillations \cite{r7}. Driving a piezoelectronic motor
at anti-resonant frequencies has also practical advantages \cite{r8}. Anti-resonance is employed to minimize unwanted
vibrations of certain parts of a system in mechanical engineering and aerospace industries. In the 
vibration control of fixture, controlling anti-resonant frequencies is more important than resonant
frequencies since the worst case can occur at anti-resonance \cite{r9,r10}. In a vibratory structural 
system an addition of mass is found to shift the resonant frequencies without affecting the anti-resonant
frequencies \cite{r11}. It has been pointed out that a shift of the resonant frequencies beyond anti-resonance
are not feasible. Anti-resonance has been realized also in quantum systems \cite{r12,r13}. 

\indent In the recent past, several studies are reported to uncover the underlying theory of anti-resonance. Particularly, stochastic
 anti-resonance is investigated in a theoretical model equation proposed for the transmission of a periodic signal mixed with a noise through static nonlinearity \cite{r34}, squid axon model equation \cite{r35}, the time evolution of interacting qubits of quantum systems \cite{r36} 
and certain piecewise linear systems \cite{r37}. Coherence anti-resonance is identified in a model of circadian rhythmicity in Drosophila 
\cite{r38} and in FitzHugh-Nagumo neuron model \cite{r39} subjected to both additive and multiplicative noise. The anti-resonance is demonstrated in a parametrically driven Van der Pol oscillator and illustrated a two state switch by using two coupled oscillators \cite{r40}. A single or 
two coupled systems are investigated in these studies with theoretical and numerical treatment. It is vital to study $n$-coupled, nonlinear 
system in the vicinity of multiple and anti-resonances.      
 
\indent Further, it is of great significance to investigate the response of coupled systems with the first unit alone subjected to external 
periodic force. Such a set up finds applications in digital sonar arrays, network of sensory neurons, 
vibrational resonance, stochastic resonance and signal propagation in coupled systems \cite{r13k,r13l,r13m,r14e,r18}. 
We report our investigation on the resonant and anti-resonant dynamics in a system of two coupled and 
$n$-coupled Duffing oscillators. The occurrence of multiple and anti-resonance are presented using
theoretical, numerical and experimental methods. The connection between the coupling strength and occurrence of multiple and anti-resonance
and, the role of system size on the $n-$ coupled system are the important significances of the present work. 
In this system, the oscillators are allowed to interact with their nearest 
neighbour with a linear coupling. Furthermore, we consider that only the first oscillator is subjected to a periodic driving
force. For a system of two-coupled oscillators and using a theoretical approach, we obtain coupled equations for the 
response amplitudes $Q_1=A_1/f$ and $Q_2=A_2/f$ where $A_1$ and $A_2$ are the amplitudes of the periodic
oscillations of the oscillator-1 and oscillator-2, respectively, and $f$ is the amplitude of the external periodic driving force. 
The theoretically predicted values of $Q_1$ and $Q_2$ are found to match very closely with the values of $Q_1$ and $Q_2$ computed 
numerically. When the frequency of the driving force is varied, two resonant peaks and one anti-resonance occur for a wide range 
of fixed values of the coupling strength $\delta$. We analyse the dependence of the resonant and anti-resonant frequencies and the
values of the response amplitudes of these two oscillators on the coupling parameter $\delta$. For the $n$-coupled systems, since theoretical analysis is very difficult and involves  solving of  $n$-coupled 
nonlinear equations for the amplitudes $A_i$'s, we analyse the occurrence of resonant and anti-resonant behaviours by 
numerically solving the system of equations of motion.  The $n$-coupled oscillators are found to show $n$ resonances 
and $n-1$ anti-resonances. We study the dependence of the resonant and anti-resonant frequencies with the number $n$ of oscillators 
that are coupled. The resonance and the anti-resonance 
behaviours found in theoretical model equations are also realized in an analog electronic circuit simulation of the 
equations. Hardware experimental analog simulation studies on two coupled and PSpice simulation of $n$-coupled oscillators shows good agreement with the theoretical/numerical predicitions.
%
\section{\label{s2}Periodically driven two-coupled systems}
The equation of motion of the $n$-coupled Duffing oscillators of our interest is 
\begin{subequations}
\label{eq1}
\begin{eqnarray} 
   \ddot{x}_1 + d \dot{x}_1 + \omega_0^2 x_1 + \beta x_1^3
                      +\delta \left( x_1 - x_2 \right)
               & = &  f \cos\omega t,   \label{eq1a}  \\
   \ddot{x}_i + d \dot{x}_i + \omega_0^2 x_i + \beta x_i^3
                      + \delta \left( x_i - x_{i-1} \right)
	              +\delta (x_{i}-x_{i+1})
               & = & 0, \label{eq1b}  \\
   \ddot{x}_n + d \dot{x}_n + \omega_0^2 x_n + \beta x_n^3
                       +\delta \left( x_n - x_{n-1} \right) 
               & = & 0.  \label{eq1c}
\end{eqnarray}
\end{subequations}
In this system the first and the last oscillators are not connected to each other and $\delta$ is the strength of 
the coupling. We start with the simplest case $n=2$, that is a system of two-coupled oscillators.
%
%
\subsection{\label{ss21}Theoretical treatment}
By applying a perturbation approach, a frequency-response equation can be obtained. We assume a periodic
solution of the system (\ref{eq1}) with $n=2$ as
\begin{eqnarray} 
    x_i(t) = a_i(t) \cos \omega t + b_i(t) \sin \omega t
 \label{eq2}
\end{eqnarray}
with $a_i$ and $b_i$ to be determined, which are slowly varying functions of time. We substitute
\begin{subequations}
 \label{eq3}
\begin{eqnarray} 
    \dot x_i(t) & = & \dot a_i \cos \omega t + \dot b_i \sin\omega t 
                    - a _i \omega \sin \omega t + b_i \omega \cos\omega t, \label{eq3a} \\
     \ddot x_i(t) & = & - 2 \dot a_i \omega \sin \omega t + 2 \dot b_i \omega \cos \omega t
                     - a_i \omega^2 \cos \omega t - b_i \omega^2 \sin \omega t, \label{eq3b} \\
     x^3_i & \approx & \frac{3}{4} \left( a_i^2 + b_i^2 \right) \left( a_i \cos \omega t 
		                 + b_i \sin\omega t \right), \label{eq3c}  
\end{eqnarray} 
\end{subequations}
where in Eq.~(\ref{eq3b}) $\ddot{a}_i$ and $\ddot{b}_i$ are neglected due to their smallness, in 
Eqs.~(\ref{eq1}) and then neglect $d\dot{a}_i$ and $d\dot{b}_i$ because they are assumed to be small.
Next, equating the coefficients of $\sin \omega t$ and $\cos \omega t$ separately to zero gives
\begin{subequations}
\label{eq4}
\begin{eqnarray}
    \dot{a}_1 & = &  \frac{b_1}{2\omega} \left[ \omega_0^2 - \omega^2+\delta
            + \frac{3 \beta }{4} \left( a_1^2 + b_1^2 \right)\right]
            - \frac{d a_1}{2}  - \frac{\delta b_2}{2 \omega},  \label{eq4a} \\
    \dot{b}_1 & = & - \frac{a_1}{2\omega} \left[ \omega_0^2 - \omega^2+\delta
            + \frac{3 \beta }{4} \left( a_1^2 + b_1^2 \right)\right]
            - \frac{d b_1}{2}  
            + \frac{\delta a_2}{2 \omega}+\frac{f}{2\omega}, \label{eq4b} \\
   \dot{a}_2 & = &  \frac{b_2}{2\omega} \left[ \omega_0^2 - \omega^2+\delta
            + \frac{3 \beta }{4} \left( a_2^2 + b_2^2 \right)\right]
            - \frac{d a_2}{2}  - \frac{\delta b_1}{2 \omega},  \label{eq4c} \\
   \dot{b}_2 & = & - \frac{a_2}{2\omega} \left[ \omega_0^2 - \omega^2+\delta
            +\frac{3 \beta }{4} \left( a_2^2 + b_2^2 \right)\right]
            - \frac{d b_2}{2}  + \frac{\delta a_1}{2 \omega}.  \label{eq4d}
\end{eqnarray}
\end{subequations}
Eqs.~(\ref{eq4}) under the transformation
\begin{equation}
\label{eq5}
    a_i(t) = A_i(t) \cos \theta_i(t), \quad b_i(t) = A_i(t) \sin \theta_i(t)
\end{equation}
take the form (with $A_i^2=a_i^2+b_i^2$)
\begin{subequations}
\label{eq6}
\begin{eqnarray}
\dot{A}_1 
        & = &  - \frac{d A_1}{2}  + \frac{\delta A_2}{2 \omega} \sin \left(\theta_{1}
		     -\theta_{2}\right) +\frac{f}{2\omega}\sin \theta_{1} = P,  
                     \label{eq6a} \\
A_1 \dot{\theta}_1 
        & = &  -\frac{A_1}{2\omega} \left[ \omega_0^2 -\omega ^2 + \delta
		+ \frac{3 \beta }{4} A_1^2 \right] 
                + \frac{\delta A_2}{2 \omega} \cos \left(\theta_{1}
		- \theta_{2}\right) + \frac{f}{2\omega}\cos \theta_{1} = Q,
                 \label{eq6b} \\
\dot{A}_2 
        & = &  - \frac{d A_2}{2}  - \frac{\delta A_1}{2 \omega} 
                    \sin \left(\theta_{1}-\theta_{2}
	             \right) = R,  \label{c11eq6c} \\
A_2 \dot{\theta}_2 
        & = & -\frac{A_2}{2\omega} \left[ \omega_0^2 - \omega ^2 
               + \delta  + \frac{3 \beta }{4} A_2^2 \right] 
               + \frac{\delta A_1}{2 \omega} \cos \left(\theta_{1}
	       - \theta_{2}\right) = S. 
\end{eqnarray}
\end{subequations}
For a periodic solution, in the long time limit, $A_{i}(t) \rightarrow A_i^*$ and 
$\theta _i(t) \rightarrow \theta_i^*$. $(A_{i}^*, \theta_i^*)$ is an equilibrium point of Eqs.~(\ref{eq6}). We set $\dot{A}_i=0$, 
$\dot{\theta}_i=0$, drop `$*$' in $A_{i}^*$ and $\theta_{i}^*$ and then eliminate $\theta_{i}$'s. We obtain the set of equations
\begin{subequations}
\label{eq7}
\begin{eqnarray}
    A_{1}^2 \left[u_{1}^2+d^{2}\omega^{2}\right] 
         + \left[2 d^{2}\omega^{2}+\delta^{2}\right]A_{2}^{2}
            - 2u_{1}u_{2}A_{2}^2  &= & f^2,  \label{eq7a} \\
    A_{2}^2 \left[u_{2}^2+d^{2}\omega^{2}\right]
      - \delta^{2} A_{1}^2 & = & 0, \label{eq7b}
\end{eqnarray}
\begin{eqnarray}
    \theta_{1} & = & \tan^{-1}\left[\frac{d \omega (A_{1}^{2}+A_{2}^{2})}{A_{1}^{2}u_{1}-A_{2}^{2}u_{2}}\right], \label{eq7c} \\
    \theta_{2} & = & \theta_{1} - \tan^{-1}\left[-\frac{d \omega}{u_2}\right], \label{eq7d}   
\end{eqnarray}
where
\begin{eqnarray}
   u_i = \omega_0^2 -\omega^2 +\delta + \frac{3 \beta}{4} A_i^2 , \quad i=1,2.
\end{eqnarray}
\end{subequations} 
The stability of the equilibrium point ($A_{1}^*, A_{2}^*, \theta_1^*, \theta_2^*$) of Eqs.~(\ref{eq6}) can be determined by 
linear stability analysis. The stability determining eigenvalues can be obtained from

\begin{equation} 
\left| \begin{array}{llll}
\dfrac{\partial P}{\partial A_1}-\lambda  & \dfrac{\partial P}{\partial \theta_1} & \dfrac{\partial P}{\partial A_2} &
 \dfrac{\partial P}{\partial \theta_2} \\
\dfrac{\partial Q}{\partial A_1} & \dfrac{\partial Q}{\partial \theta_1} - \lambda & \dfrac{\partial Q}{\partial A_2} &
 \dfrac{\partial Q}{\partial \theta_2} \\
\dfrac{\partial R}{\partial A_1} & \dfrac{\partial R}{\partial \theta_1} & \dfrac{\partial R}{\partial A_2} -\lambda &
 \dfrac{\partial R}{\partial \theta_2} \\
\dfrac{\partial S}{\partial A_1} & \dfrac{\partial S}{\partial \theta_1} & \dfrac{\partial S}{\partial A_2} &
 \dfrac{\partial S}{\partial \theta_2} -\lambda \end{array} \right|=0,
\end{equation}
Where the partial derivatives are evaluated at the equilibrium point. Expanding the determinant we obtain the characteristic equation of the
form
\begin{equation} 
a_4\lambda^{4} + a_3\lambda^{3} + a_2\lambda^{2} + a_1\lambda + a_0 = 0. \label {cheq}
\end{equation}

The equilibrium point ($A_{1}^*, A_{2}^*, \theta_1^*, \theta_2^*$) is stable if all the eigenvalues of Eq.~(\ref{cheq}) has 
negative real part; otherwise it is unstable. For a stable ($A_{1}^*, A_{2}^*, \theta_1^*, \theta_2^*$) the system-(\ref{eq1}) exhibits
a stable periodic solution.   
 
%
\subsection{\label{ss22}Two-coupled linear systems}
We consider now the case of two-coupled linear and undamped oscillators ($n=2$, $\beta=0$ and $d=0$ in Eqs.~(\ref{eq1}))
\cite{r1} whose amplitudes $A_1$ and $A_2$ which are obtained from Eqs.~(\ref{eq7}) are 
\begin{subequations}
\label{eq8}
\begin{eqnarray}
      A_1 & = & \frac{f(\omega_{0}^2 - \omega^2 + \delta)}{(\omega_{0}^2
                    - \omega^2)(\omega_{0}^2-\omega^2+2\delta)}, 
		      \label{eq8a} \\ 
      A_2 & = &  \frac{\delta A_{1}}{(\omega_{0}^2-\omega^2+\delta)}
                        \nonumber \\ 
            & = &  \frac{\delta f}{(\omega_{0}^2-\omega^2)(\omega_{0}^2
		  - \omega^2+2\delta)}.\label{eq8b} 
\end{eqnarray}
\end{subequations} 
%
%
\begin{figure}[!t]
\centering
\includegraphics[width=0.8\columnwidth]{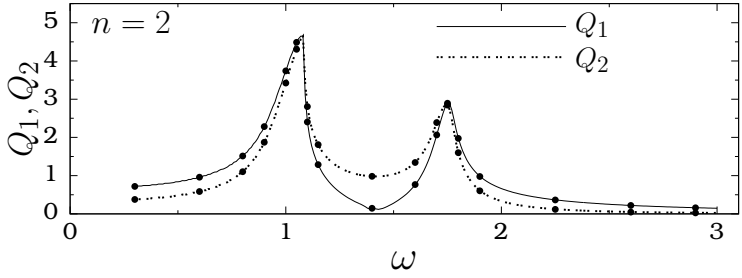}%
\caption{Response amplitudes $Q_1$ (of oscillator-1) and $Q_2$ (of oscillator-2) versus the frequency $\omega$ of the
driving force of the two-coupled Duffing oscillators, Eqs. (\ref{eq1}) with $n=2$. The continuous and dotted curves 
are the theoretically predicted $Q_1$ and $Q_2$, respectively. The solid circles are the numerical ones.  Here $d=0.1$, 
$\omega^{2}_{0}=1$, $\beta=1$, $\delta=1$ and $f=0.1$}%
\label{fig1}%
\end{figure}
%
Both $A_{1}$ and $A_{2}$ are a maximum at $\omega=\omega_{0}$ and $\sqrt{\omega_{0}^2+2\delta}$. 
The amplitude $A_{1}$ becomes minimum when the term $\left( \omega_{0}^2-\omega^2+\delta \right)$ in Eq.~(\ref{eq8a})
is a minimum. This happens at $\omega=\sqrt{\omega_{0}^2+\delta}$. Thus, the frequency at which 
anti-resonance occurs in oscillator-1 is $\omega_{1,{\mathrm{ar}}}=\sqrt{\omega_{0}^2+\delta}$. 
At $\omega_{1,{\mathrm{ar}}}$, $A_{1,{\mathrm{ar}}}=0$. From Eq.~(\ref{eq8b}), it is reasonable to 
expect $A_{2}$ to be a minimum at a frequency $\omega$ at which $A_{1}$ becomes a minimum. Substitution of 
$\omega^2=\omega^2_{1,{\mathrm{ar}}}=\omega^2_0+\delta$ in Eq.~(\ref{eq8b}) gives
$A_{2,{\mathrm{ar}}}=f/\delta \neq A_{1,{\mathrm{ar}}}$ and is nonzero. 

For the damped $(d \neq 0)$ and linear system we have
\begin{subequations}
\label{eq9}
\begin{eqnarray}
     A_1 
        & = &  \left(\frac{f^2 u_{+}}{u^{2}_{+}+\delta^4
                  -2\delta^2 u_{-}}\right)^{1/2}, \label{eq9a} \\
     A_2 
        & = &  \frac{\delta A_{1}}{\sqrt{u_{+}}}, 
                \quad u_{\pm}=(\omega_{0}^2-\omega^2+\delta)^2 
                \pm d^2\omega^2. \label{eq9b}
\end{eqnarray}
\end{subequations} 
It is difficult to obtain explicit expressions for the two resonant frequencies and the corresponding amplitudes
due to the complexity of the expressions of $A_{1}$ and $A_{2}$. However, the anti-resonant frequency can be determined by
seeking the value of $\omega$ at which the quantity $u_{+}$ becomes a minimum. This gives
\begin{equation}
\label{eq10}
 \omega_{1,{\mathrm{ar}}} = \sqrt{\omega_{0}^2
               +\delta-\frac{d^2}{2}}.
\end{equation}

\subsection{\label{ss23}Two-coupled Duffing oscillators}
Now, we consider the two-coupled Duffing oscillators. Decoupling of the amplitudes in Eqs.~(\ref{eq7}) is very 
difficult. However, applying the Newton-Raphson method \cite{r14} developed for coupled equations, various possible
values of $A_{1}$ and $A_{2}$ can be determined and then the frequency-response curve can be drawn.
%
\begin{figure}[!ht]
\centering
\includegraphics[width=0.80\columnwidth]{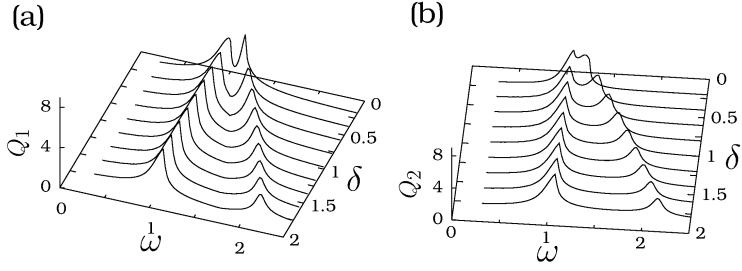}
\caption{Response amplitudes ({\bf a}) $Q_1$ and ({\bf b}) $Q_2$ as a function of the parameters $\delta$ and
$\omega$ for the system (1) with $n=2$, $d=0.1$, $\omega^{2}_{0}=1$, $\delta=1$ and $f=0.1$}%
\label{fig2}
\end{figure}
%

We choose the values of the parameters as $d=0.1, \omega_{0}^2=1, \beta=1$ and $\delta=1$. Figure~\ref{fig1} 
presents both theoretical $Q_{i} (=A_{i}/f)$ and numerically computed $Q_{i}$, $i=1,2$ as a function of 
the driving frequency $\omega$. The theoretical prediction matches very closely the numerical results obtained from the 
simulations. $Q_{1}$ is a maximum at $\omega=1.08$ and also at $\omega=1.75$. $Q_2$ becomes a maximum at $\omega=1.08$ 
and at $1.74$. Anti-resonance in $Q_{1}$ and $Q_{2}$ occurs at $\omega_{1,{\mathrm{ar}}}=1.43$ and $\omega_{2,{\mathrm{ar}}}=1.42$. 
However, $Q_{1,{\mathrm{ar}}}=0.1181$ while $Q_{2,{\mathrm{ar}}}=0.9804$. In order to show the significant effects of the linear 
coupling constant $\delta$ on the resonant dynamics, we display the dependence of $Q_{1}$ and $Q_{2}$ versus $\omega$ on the 
parameter $\delta$ in Fig.~\ref{fig2}. $Q_{1}$ has a single resonance for $0<\delta<0.1$ and two resonant peaks for $\delta\geq0.1$. For the second oscillator a double resonance is realized for $\delta\geq0.13$.

\begin{figure}[ht]
\centering
\includegraphics[width=0.85\columnwidth]{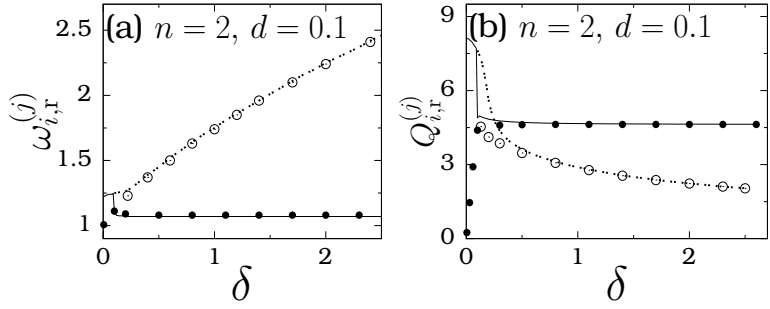}
\caption{Dependence of ({\bf a}) resonant frequencies and ({\bf b}) the response amplitudes at the resonant frequencies as a 
function of the coupling constant $\delta$ of the two-coupled Duffing oscillators. In both subplots, the continuous and 
dotted lines are associated with the first and second resonances of the first oscillator, respectively. The solid and open 
circles correspond to the first and second resonances of the second oscillator, respectively. Here $d=0.1$, $\omega^{2}_{0}=1$, 
$\beta=1$ and $f=0.1$}
\label{fig3}
\end{figure}

We denote $\omega_{i,{\mathrm{r}}}^{(j)}$ as the value of $\omega$ at which the $i$th resonance occurs in the $j$th 
oscillator. And $Q_{i,{\mathrm{r}}}^{(j)}$ is the value of the response amplitude at $\omega_{i,{\mathrm{r}}}^{(j)}$. 
For $n=2$ there is only one anti-resonance. Therefore, we denote $\omega_{1,{\mathrm{ar}}}$ and 
$\omega_{2,{\mathrm{ar}}}$ as the values of $\omega$ at which anti-resonance occurs in the oscillators 1 and 2, 
respectively, and the corresponding values of $Q$ as $Q_{1,{\mathrm{ar}}}$ and $Q_{2,{\mathrm{ar}}}$, 
respectively. These quantities are computed for a range of values of $\delta$. Figure~\ref{fig3}(a) 
displays the variation of $\omega_{1,{\mathrm{r}}}^{(1)}$ (continuous curve), 
$\omega_{2,{\mathrm{r}}}^{(1)}$ (dotted curve), $\omega_{1,{\mathrm{r}}}^{(2)}$ (solid circles) and 
$\omega_{2,{\mathrm{r}}}^{(2)}$ (open circles) with $\delta$. The first resonant frequencies of both oscillators are almost the 
same and independent of $\delta$, except for $\delta \ll 1$. In contrast to this, the second resonant frequencies of the two 
oscillators vary with $\delta$,  however, 
$\omega_{2,{\mathrm{r}}}^{(1)} \approx \omega_{2,{\mathrm{r}}}^{(2)}$ for each fixed value of $\delta$, except for 
$\delta \ll 1$. In Fig.~\ref{fig3}(b) $Q^{(j)}_{1,{\mathrm{r}}}$, $j=1,2$ approach the same constant value where as 
$Q^{(j)}_{2,{\mathrm{r}}}$, $j=1,2$ decreases for increasing values of $\delta$.

\begin{figure}[t]
\centering
\includegraphics[width=0.95\columnwidth]{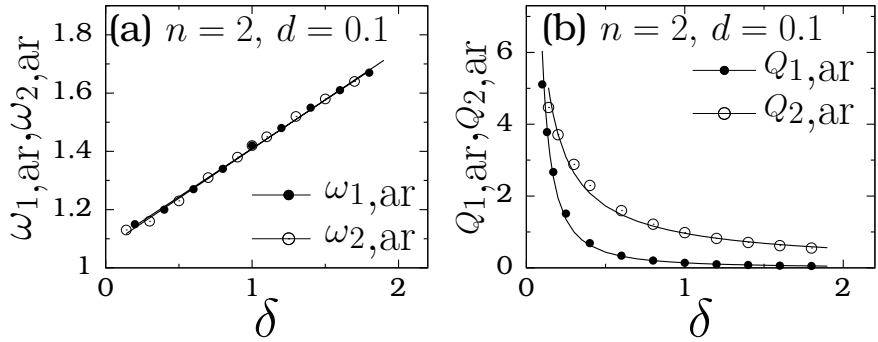}
\caption{Variation of ({\bf a}) anti-resonant frequencies $\omega_{1},_{\mathrm{ar}}$ and $\omega_{2},_{\mathrm{ar}}$ 
of the first and second oscillators, respectively, and ({\bf b}) response amplitude of the anti-resonance with the coupling 
strength $\delta$ of the two-coupled Duffing oscillators, Eqs.~(\ref{eq1}) with $n=2$. The symbols are numerical data and 
continuous curves are the best fit.}
\label{fig4}
\end{figure}

The dependence of anti-resonance frequencies and the corresponding amplitudes of the oscillations on the coupling 
strength $\delta$ are plotted in Fig.~\ref{fig4}. 
In this figure the numerical results are represented by symbols and the appropriate curve fits are 
marked by continuous curves. Furthermore, the frequencies $\omega_{1,{\mathrm{ar}}}$ and $\omega_{2,{\mathrm{ar}}}$ are found 
to depend linearly on $\delta$. We obtain $\omega_{1,{\mathrm{ar}}} = 1.0759 + 0.325 \delta$ and 
$\omega_{2,{\mathrm{ar}}} = 1.079 + 0.325\delta$. In the linear system ($\beta=0$), as noted earlier, 
$\omega_{1,{\mathrm{ar}}}=\sqrt{\omega_{0}^2+\delta}\approx 1 + 0.5\delta$. $Q_{1,{\mathrm{ar}}}$ and 
$Q_{2,{\mathrm{ar}}}$ decreases rapidly following the power-law relations $0.14106 \delta^{-1.696}$ 
(for $\delta >0.1$) and $0.939 \delta^{-0.881}$ (for $\delta>0.13$), respectively. That is, by increasing the value of $\delta$, 
the anti-resonant frequency is increased, while the response amplitude at the anti-resonance is reduced. In 
Fig.~\ref{fig4} $\omega_{1,{\mathrm{ar}}}^{(1)} \approx \omega_{2,{\mathrm{ar}}}^{(2)}$ but 
$Q_{1,{\mathrm{ar}}} < Q_{2,{\mathrm{ar}}}$ for a wide range of values of $\delta$. By increasing the value of 
the damping coefficient, the response amplitudes decrease and for sufficiently large values, the second resonance in both oscillators
 is suppressed.
%
%
\begin{figure}[h!]
\centering
\includegraphics[width=0.9\columnwidth]{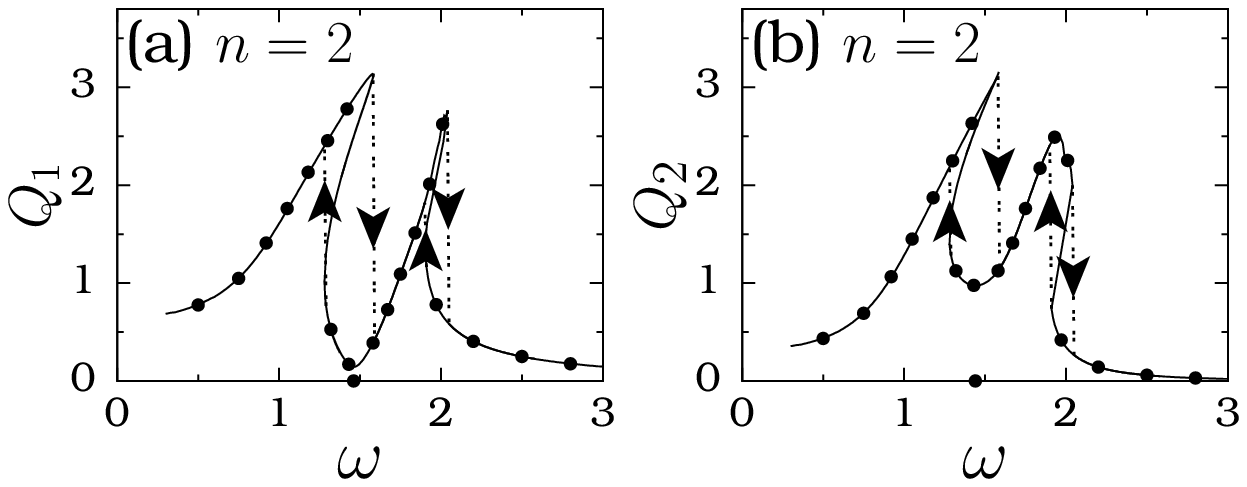}
\caption{Frequency-response curves of ({\bf a}) oscillator-1 and ({\bf b}) oscillator-2, of the two-coupled 
Duffing oscillators (Eqs.~(\ref{eq1})) with $n=2$. The continuous curve and the solid circles are 
the theoretically predicted and numerically computed values of response amplitudes. 
The solid circle on the $\omega$ axis mark the value of $\omega$ at which anti-resonance occurs. 
In both subplots the downward and upward arrows indicate the jump in the response amplitudes when the frequency is varied 
in the forward and backward directions, respectively. 
The values of the parameters are $d=0.1$, $\omega^{2}_{0}=1$, $\beta=20$, $\delta=1$ and $f=0.1$}
\label{fig5}
\end{figure}

One of the features of the resonance in nonlinear oscillators is the appearance of hysteresis in the frequency-response
curve. This is also realized in the system (\ref{eq1}) near two resonances for certain range of values of 
$\beta$. An example is presented in Fig.~\ref{fig5} for $\beta=20$ and $\delta=1$. In both oscillators two stable 
periodic orbits with different amplitudes coexist for $\omega \in [1.28,1.58]$ and $[1.9,2.04]$. In these intervals, the theoretical response curve has three branches. The upper branch and lower branch are realized when the frequency
is swept in the forward and backward directions, respectively. These are stable branches and are observed in the 
numerical simulations, as well. For each $\omega$ in the intervals $[1.28, 1.58]$  and $[1.9,2.04]$ there exist two 
periodic orbits with different amplitudes. They are observed for different set of initial conditions. 
The middle branch is not realized in the numerical simulations for a large set of initial conditions and is an unstable branch.
When $\omega$ is increased from a small value, a first resonance in both oscillators occurs at $\omega=1.58$ but 
with $Q_{1}=3.14$ and $Q_{2}=3.11$. The second resonance in the oscillators 1 and 2 takes place at $\omega=2.04$ 
and $1.95$, respectively, with $Q_{1}=2.76$ and $Q_{2}=2.52$.

\begin{figure}[tbh]
\centering
\includegraphics[width=0.7\columnwidth]{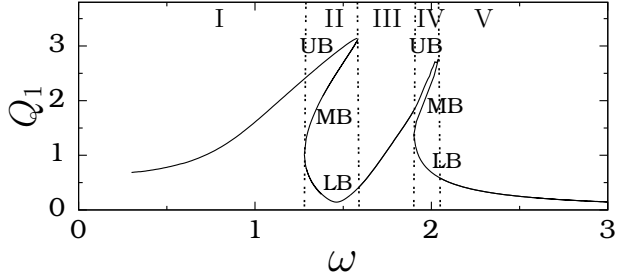}
\caption{Frequency-response curve of oscillator-1. The intervals I, II, and V have only one stable periodic orbit and the 
remaining intervals have three branches in which the upper branch (UB) and lower branch (LB) correspond to stable periodic orbits,
while the middle branch (MB) correspond to a unstable periodic orbit. 
The values of the parameters are $d=0.1$, $\omega^{2}_{0}=1$, $\beta=20$, $\delta=1$ and $f=0.1$}
\label{fig5a}
\end{figure}

The stability analysis of equilibrium points of Eqs.~\ref{eq6} is performed to determine the stability of the periodic orbits of the 
system~(\ref{eq1}). From Eqs.~(\ref{eq7a}) and (\ref{eq7b}), $A_i$s are calculated which are then substituted in Eqs.~(\ref{eq7c}) and 
(\ref{eq7d}) to get $\theta_i$'s. These ($A_{1}^*, A_{2}^*, \theta_1^*, \theta_2^*$)s calculated for various values of $\omega$ is 
applied to Eq.(\ref{cheq}) to obtain the eigenvalues of the characteristic equation. The frequency response curve is classified into 
five segments as shown in Fig.~\ref{fig5a}. The odd numbered segments are have only one ($A_{1}^*, A_{2}^*, \theta_1^*, \theta_2^*$).
Complex conjugate eigenvalues with negative real part are obtained in these regions. This implies that 
($A_{1}^*, A_{2}^*, \theta_1^*, \theta_2^*$) is stable in these region. Hence, the periodic orbit in these 
regions are stable. In the even numbered intervals of $\omega$, there are three equilibrium points and are classified into three branches.
The upper and lower branches (UB and LB) have the complex conjugate eigenvalues with negative real part while the middle branch (MB) 
has an eigenvalue with positive real part. Hence, the UB and LB are stable while the MB is unstable. 
Table-1 presents stability determining eigenvalues for certain specific values of $\omega$. 
\vspace{-0.5cm}

\begin{table}[]
\centering
\label{t1}
\begin{tabular}{llllllll}
\hline
\multicolumn{1}{c}{Region in $\omega$} & \multicolumn{1}{c}{$\omega$} & \multicolumn{1}{c}{$A_{1}^{*}$} & \multicolumn{1}{c}{$A_{2}^{*}$} & 
\multicolumn{1}{c}{$\theta_{1}^{*}$} & \multicolumn{1}{c}{$\theta_{2}^{*}$} & \multicolumn{1}{c}{Eigenvalues} & \multicolumn{1}{c}{Nature of Stability} \\ \hline
    I       & 1.0  & 0.16076  &  0.12828       &  0.26899      &  0.34902    & $-0.05024 \pm {\mathrm{i}}1.31391$,  & Stable   \\ 
& & &  & & & $-0.05019 \pm {\mathrm{i}}0.27370$ \\ \hline
     & 1.35 & 0.25929  &  0.24097       &  0.73131      &  0.85937     &  
$-0.05060 \pm {\mathrm{i}}1.08736$, & Stable   \\
& & &  & & & $-0.05057 \pm {\mathrm{i}}0.19524$ \\ 
&&&&&&\\ 
    II  & 1.35 & 0.18618  &  0.21345       &  0.62381      &  -0.46827    &
$-0.13278 \pm {\mathrm{i}}0.70677$, & Unstable  \\
& & &  & & & $-0.11951$, $0.07871$ \\
&&&&&&\\
     & 1.35 & 0.03760  &  0.10497       &  0.44762      &   -0.07242  &                     
 $-0.47278 \pm {\mathrm{i}}1.81332$ & Stable                     \\ 
& & &  & & & $-0.01427 \pm {\mathrm{i}}0.07318$ \\ \hline
    III     & 1.75 & 0.10911   &  0.17609        &  0.75379      &   0.46881    & 
$-0.14187 \pm {\mathrm{i}}0.81336$,  & Stable   \\
& & &  & & & $-0.03768 \pm {\mathrm{i}}0.10825$ \\ \hline
    & 2.0 &  0.25589  &  0.22688       &    -1.42475      &   -1.58621      & 
$-0.01502 \pm {\mathrm{i}}0.17607$, & Stable   \\
& & &  & & & $-0.01494 \pm {\mathrm{i}}3.77292$\\
&&&&&&\\
    IV & 2.015 & 0.24531   &  0.20141        &   -1.08309        &   -1.22100                  &
$~~~0.00065 \pm {\mathrm{i}}0.68099$, & Unstable   \\ 
& & &  & & & $~~~0.00080 \pm {\mathrm{i}}0.08891$\\
&&&&&&\\
    & 2.0 &  0.06951  & 0.03524  &   -0.17544           &    -0.27604          &    
$-0.00011 \pm {\mathrm{i}}0.24151$,  & Stable   \\
& & &  & & & $-0.00011 \pm {\mathrm{i}}0.74499$\\ \hline
    V       & 2.5  &     0.02513    & 0.00633         &  -0.06684       &  -0.12560     &                  
$-0.00092 \pm {\mathrm{i}}0.61578$, & Stable   \\ 
& & &  & & & $-0.00068 \pm {\mathrm{i}}1.02009$\\ \hline
\end{tabular}
\caption{The equilibrium points and their stability determining eigenvalues for certain values of $\omega$ corresponding to Fig.~\ref{fig5a}.} 
\end{table}

\newpage
\subsection{\label{ss24}Analog simulation}
\begin{figure}[h!]
\centering
\includegraphics[width=0.65\columnwidth]{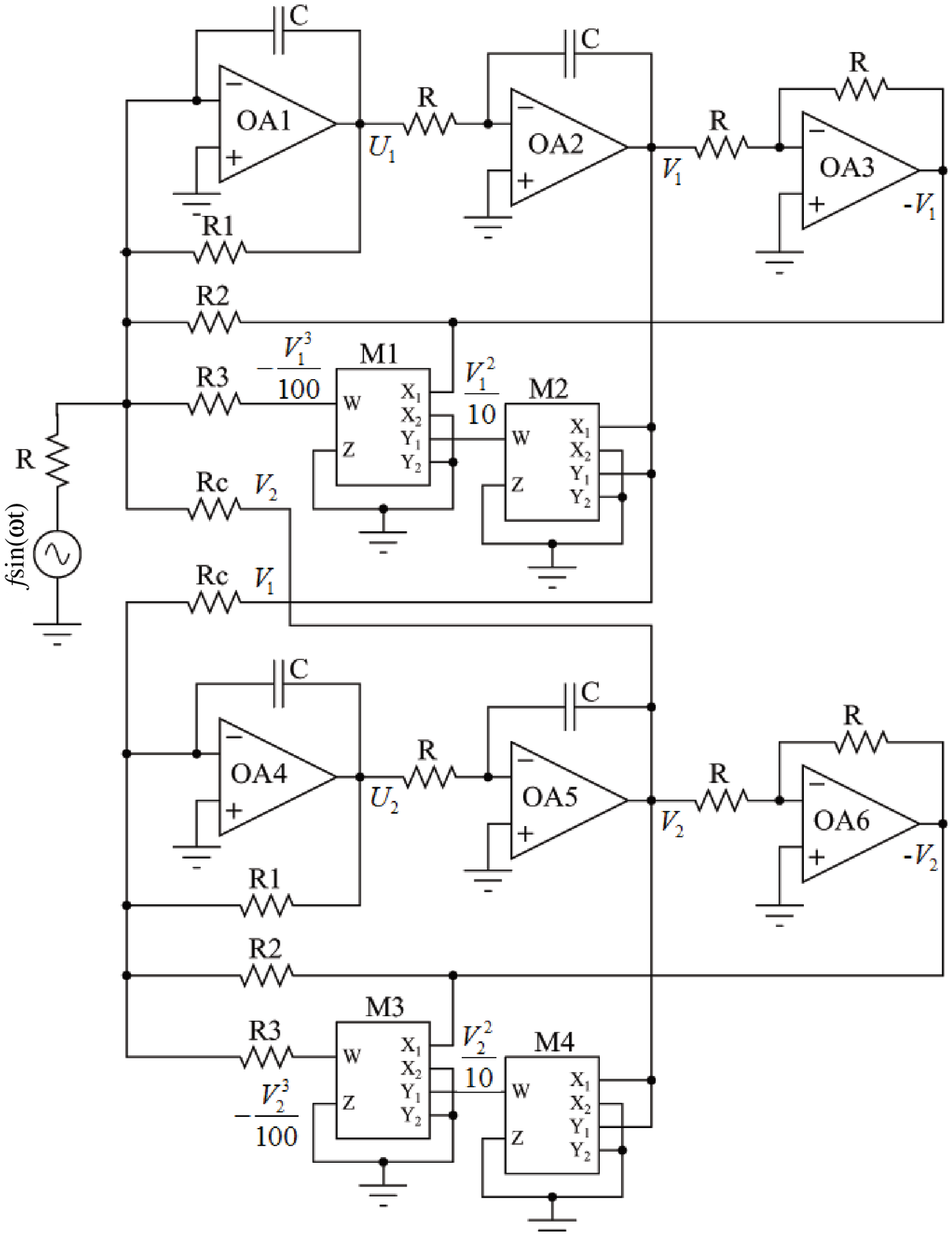}
\caption{The analog circuit for the two-coupled Duffing oscillators. `OA's are TL082 op-amps and `M's are AD633JN multiplier ICs}
\label{fig6}
\end{figure}

The multiple resonance and anti-resonance found in the theoretical and numerical studies of the Duffing oscillator 
system (\ref{eq1}) can be realized in the analog electronic circuit simulation. Figure \ref{fig6} presents the analog 
circuit for the Eq.~(\ref{eq1}) with $n=2$.  The evolution equations for the variables $V_1$ and $V_2$ obtained 
using the Kirchhoff's voltage law are
\begin{subequations}
\label{eq11}
\begin{eqnarray}
	R^2 C^2 \frac{d^2 V_1}{dt^2} 
        & = & -\left(\frac{R^2C}{R1}\right)\frac{dV_1}{dt}
               - \left(\frac{R}{R2}\right)V_1
	       - \left(\frac{R}{100R3}\right)V^{3}_{1} \nonumber \\
        & &    \quad   + \left(\frac{R}{R_C}\right)V_2 + f \sin \omega t,
                \label{eq11a} \\
	R^2 C^2 \frac{d^2 V_2}{dt^2}
        & = & -\left(\frac{R^2C}{R1}\right)\frac{dV_2}{dt} 
              - \left(\frac{R}{R2}\right)V_2
	      - \left(\frac{R}{100R3}\right)V^{3}_{2} 
              + \left(\frac{R}{R_C}\right)V_1. \label{eq11b}
\end{eqnarray}
\end{subequations} 

In order to bring Eqs.~(\ref{eq11}) in dimensionless form, we introduce the change of variables $t/RC=t'$, 
$\omega/RC=\omega'$, $V_1=x_1$ and  $V_2=x_2$ and then drop the primes in $t'$ and $\omega'$. The result 
is the following set of equations,
\begin{subequations}
\label{eq12}
\begin{eqnarray}
	\ddot{x}_1 + d \dot{x}_1 + (\omega^{2}_{0} + \delta)x_1 + \beta x^{3}_1 - \delta x_2 
	           & = & f \sin \omega t, 	\label{eq12a} \\
	\ddot{x}_2 + d \dot{x}_2 + (\omega^{2}_{0} + \delta)x_2 + \beta x^{3}_2 - \delta x_1 & = & 0, \label{eq12b}
\end{eqnarray}
\end{subequations} 
where $d=R/R1$, ${\mathrm(}\omega^{2}_{0} + \delta {\mathrm)} =R/R2$, $\beta=R/R3$, and $\delta=R/R_C$. 
In Fig.~\ref{fig6}, the values of $R$ and $C$ are fixed as $10~$k${\mathrm{\Omega}}$ and  $96.4~$nF, respectively.
The values of the resistors $R1$, $R2$, $R3$ and $R_C$ can be varied to change the values of $d$, $\omega^{2}_{0}$, 
$\beta$ and $\delta$, respectively. We performed the analog circuit simulation of two-coupled Duffing oscillators 
using the circuits implemented on bread board. The components in the circuit are carefully chosen with less than 
$1\%$ tolerance in the values. To obtain the response amplitudes ($A_1~\&~A_2$), fast Fourier transform (FFT) analysis 
on the output of each oscillator is performed using mixed signal oscilloscope Agilent MSO6014A. Small fluctuations 
are noted in $A_i$s observed in the FFT displayed in the scope. For the better accuracy an average value of $A_i$s over 
$10$ measurements is obtained. The values of $A_i$s  are displayed in dBV. It is converted into units of volt (V) 
using the relation ${\mathrm{dB=20log\left(\frac{V}{V_0}\right)}}$ with ${\mathrm{V_0=1.0V}}$. In the analog simulation, 
we fixed the circuit component values in equivalence with the values of the parameters used in the 
theoretical/numerical study. 
\begin{figure}[!t]
\centering
\includegraphics[width=0.8\columnwidth]{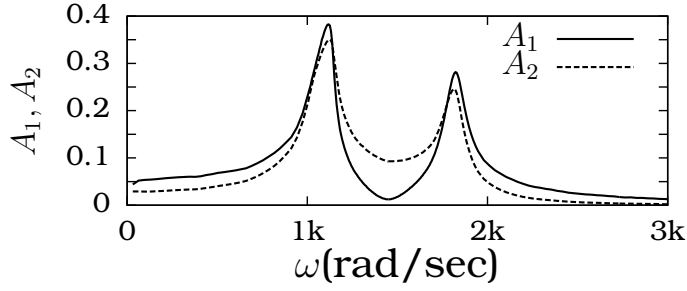}
\caption{Experimentally obtained response amplitude $A_1$ (oscillator-1) and $A_2$ (oscillator-2) versus the frequency $\omega$ 
of the input periodic signal. Here, $d=0.1$ ($R1=100~{\mathrm{k\Omega}}$), $\omega^{2}_{0}=1$, $\beta=1$ 
($R3=10~{\mathrm{k\Omega}}$), $\delta=1$ ($R2=5~{\mathrm{k\Omega}}$, $R_C=10~{\mathrm{k\Omega}}$) and $f=0.1$(in V)  }
\label{fig7}
\end{figure}

Figure \ref{fig7} shows response amplitudes $A_1$ and $A_2$ (in units of volts) versus $\omega$ of the two-coupled 
Duffing oscillators circuits. The values of the parameters are as in Fig.~\ref{fig1}.
$A_1$ ($A_2$) is maximum at  $\omega=\omega_{\mathrm{r}}=1.12~{\mathrm{krad/sec}}$, $1.82~{\mathrm{krad/sec}}$
$(1.12~{\mathrm{krad/sec}}, 1.815~{\mathrm{krad/sec}})$. Anti-resonance in $A_1$ and $A_2$ occurs at
$\omega_1,_{\mathrm{ar}}=1.45~{\mathrm{krad/sec}}$ and $\omega_2,_{\mathrm{ar}}=1.45~{\mathrm{krad/sec}}$. 
These values agree very closely with the theoretical/numerical predictions. A small deviations observed in the 
experimental values with theory/numeric are listed in table-2.

\begin{table}[h]
\centering
\label{t2}
\begin{tabular}{lcccccc}
\hline
\multicolumn{1}{c}{} & $\omega^{(1)}_{1,{\mathrm{r}}}$ & $\omega^{(1)}_{2,{\mathrm{r}}}$ & $\omega^{(2)}_{1,{\mathrm{r}}}$ & $\omega^{(2)}_{2,{\mathrm{r}}}$ & $\omega^{(1)}_{\mathrm{ar}}$ & $\omega^{(2)}_{\mathrm{ar}}$ \\ \hline
                    Theory & 1.08 & 1.75 & 1.08 & 1.74 & 1.43 & 1.42 \\
                     Experiment & \multirow{2}{*}{} & \multirow{2}{*}{} & \multirow{2}{*}{} & \multirow{2}{*}{} & \multirow{2}{*}{} & \multirow{2}{*}{} \\
(krad/Sec) &         1.12          &        1.82           &         1.12          &        1.815           &        1.45           &     1.45              \\ \hline
\end{tabular}
\caption{Comparison of theoretically and experimentally computed values of $\omega^{(j)}_{i,{\mathrm{r}}}$ and 
$\omega^{(j)}_{{\mathrm{ar}}}$ for the two-coupled Duffing oscillators.}
\end{table}

\begin{figure}[!b]
\centering
\includegraphics[width=0.95\columnwidth]{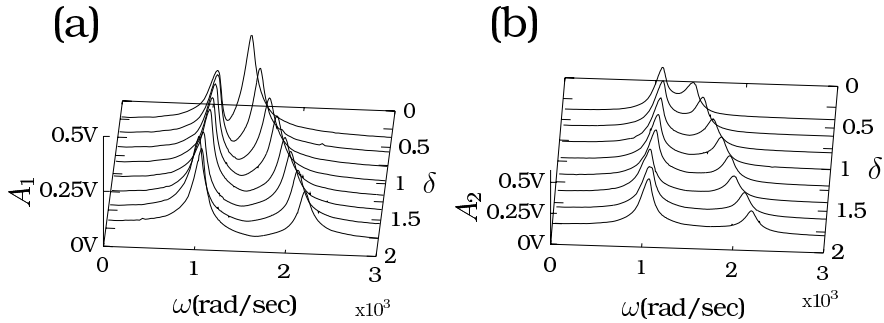}
\caption{Experimentally obtained response amplitudes $A_1$ and $A_2$ as a function of the parameters $\delta$ and $\omega$ for the
system (\ref{eq1}) with $n=2$, $d=0.1$, $\omega^{2}_{0}=1$, $\beta=1$, $\delta=1$ and $f=0.1$}
\label{fig8}
\end{figure}

The effect of the coupling strength $\delta$, on the oscillator-1 and oscillator-2 of the two-coupled Duffing 
oscillators is shown in Figs.~\ref{fig8}(a) \&  (b), respectively. In Figs. \ref{fig8} we can clearly notice 
that the second resonance frequency, and the frequency of anti-resonance, of both oscillators increase with 
increase in the value of $\delta$, while the first resonant frequency of the two oscillators settles to a constant value. 
The results are in agreement with the theoretical prediction. It is noted that the multiple and anti-resonance are observed for 
small strength of nonlinearity ($\beta$). For a large value of $\beta$, the corresponding controlling resistor ($R3$) in the analog circuit
becomes very small. This creates the impedance mismatch in the input stage of the first integrator of the each oscillator in the coupled
circuit which eventually brings down the operation of the circuit. Due to that it is difficult to obtain hysteric frequency response 
curve in the analog circuit of present configuration. 
%
\section{\label{s3}Response of the system of $n$-coupled oscillators}
In this section, we report our investigation on system of $n-$coupled Duffing oscillators. 
For the case of $n(>2)$-coupled oscillators and by following the theoretical procedure employed for the $n=2$ case a set of 
$n$-coupled nonlinear equations for the amplitudes $A_{i}$ can be obtained. Solving them analytically or numerically is very 
difficult. Therefore, we analyse the case of $n>2$ by numerically integrating the Eqs.~(\ref{eq1}) and computing
the amplitudes $A_{i}$'s and then the response amplitudes $Q_{i}$'s.

\begin{figure}[t]
\centering
\includegraphics[width=1.0\columnwidth]{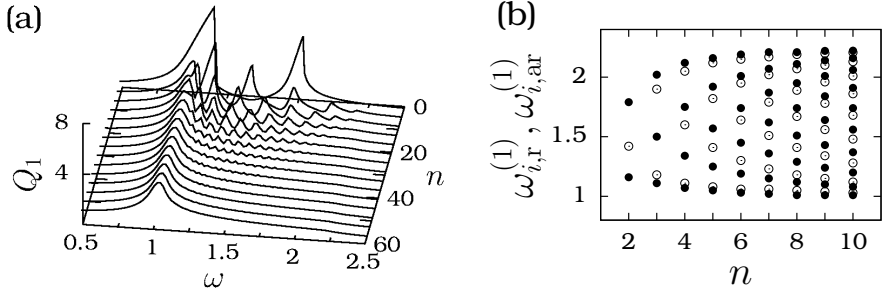}
\caption{({\bf a}) Frequency-response curve of the first oscillator as a function of $n$ (number of oscillators coupled) 
in the system (\ref{eq1}) for some selective values of $n$ in the interval [2,60]. ({\bf b}) Variation of the resonance 
frequencies $\omega_{i,{\mathrm{r}}}^{(1)}$, $i=1,2,\cdots,n$ (solid circles), and the anti-resonant frequencies $\omega_{i,
{\mathrm{ar}}}^{(1)}$, $i=1,2,\cdots,n-1$ (open circles) of the first oscillator as a function of the number of oscillator $n$. 
The values of the parameters are $d=0.05$, $\omega^{2}_{0}=1$, $\beta=1$, $\delta=1$ and $f=0.1$}
\label{fig9}
\end{figure}

We fix the values of the parameters as $d=0.05$, $f=0.1$, $\omega_{0}^2=1$, $\beta=1$ and $\delta=1$. 
Figure~\ref{fig9} presents $Q_{1}$ versus $\omega$ for $n=2,3,\cdots,60$. In Fig.~\ref{fig9}(a) for the first values of 
$n$, the frequency-response curve displays clearly $n$ distinct resonant peaks and $n-1$ anti-resonances (minimum values 
of the response amplitude). The response amplitude at successive resonances in each oscillator generally decreases. 
For, say, $n<10$, the last resonance peak is visible. For sufficiently large values of $n$, the resonance suppression 
and reduction in the response amplitude takes place. 

For the first oscillator, we numerically compute the values of the $n$ resonant frequencies 
$\omega_{i,{\mathrm{r}}}^{(1)}$, $i=1,2,\cdots,n$ at which the response amplitude becomes maximum
and the $n-1$ anti-resonant frequencies $\omega_{i,{\mathrm{ar}}}^{(1)}$, $i=1,2,\cdots,n-1$ at which
the response amplitude becomes locally minimum. The result is presented in Fig.~\ref{fig9}(b) for 
$n\in[2,10]$.  For $n=2$ and $n=10$ there are $2$ and $10$ resonances, respectively, and $1$ and $9$ anti-resonances, 
respectively. These are clearly seen in Fig~\ref{fig9}(b). A remarkable result in Fig.~\ref{fig9} is that as 
$n$ increases from $2$ the values of the first resonant frequency $\omega_{1,{\mathrm{r}}}^{(1)}$, 
the response amplitude $Q_{1,{\mathrm{r}}}^{(1)}$ at the first resonant and first anti-resonant frequency 
$\omega_{1,{\mathrm{ar}}}^{(1)}$ decreases and approach a limiting value.
For the chosen parametric values, the limiting value of $\omega_{1,{\mathrm{r}}}^{(1)}$ is $1.04$, which is a value close to the 
natural frequency $\omega_0=1$ of the uncoupled linear oscillators.
The limiting value of $Q_{1,{\mathrm{r}}}^{(1)}$ is $\approx3.38$. The value of $\omega$ at which the last resonance takes place 
(denoted as $\omega_{n,{\mathrm{r}}}^{(1)}$) increases with $n$ and attains a saturation at $2.23$.
The response amplitude at the last resonance decreases with $n$. We denote $\omega'$ and $\omega''$ as the
limiting values of the resonant frequencies of the first and the last resonance, respectively. Then, as $n$ increases the 
newer and newer resonant and anti-resonant frequencies should fall within the frequency interval $[\omega',\omega'']$ with a 
decreasing response amplitude at successive resonances. Essentially, the resonance profile displays an amplitude modulation. 
In Fig.~\ref{fig9}(a) such modulation is visible for $n \in [20,40]$. The modulation is weak for
sufficiently large values of $n$ as is the case of $n=60$, where the frequency-response curve shows a single 
resonant peak. 

The response of the system of $n$-coupled Duffing oscillators with different types of linear coupling is numerically investigated.
Multiple resonances and anti-resonances are found to occur in a system of small number of coupled oscillators with
a coupling of the form $\delta (x_{i}-x_{i-1})+\delta (x_{i}-x_{i+1})$. For the coupling of the form 
$\delta (\dot x_{i}-\dot x_{i-1})+\delta (\dot x_{i}-\dot x_{i+1})$ multiple resonances and anti-resonances are 
not found. These resonances are observed in the case of a coupling of the form $\delta (x_{i-1}+x_{i+1})$. When all the 
oscillators are driven by periodic forces, then only a single resonance is obtained for the different kinds of 
coupling considered in the present work. It is to be remarked that the resonant behaviours observed in the system (\ref{eq1}) are not 
realized in unidirectionally coupled Duffing oscillators \cite{r18}.

\begin{figure}
\centering
\includegraphics[width=0.65\columnwidth]{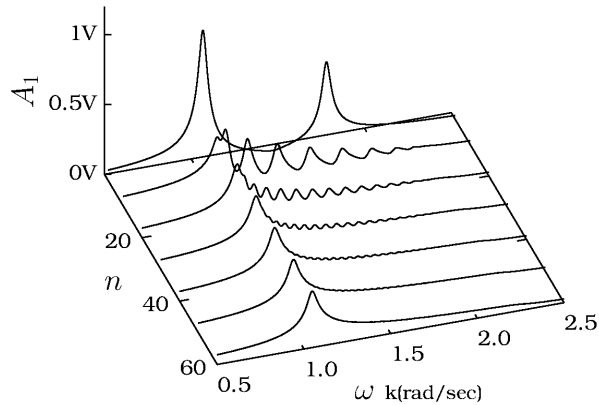}%
\caption{Experimental frequency-response curve of the first oscillator as a function of $n$ in the circuit corresponding
to the system (\ref{eq1}) for a few selective values of $n$}
\label{fig10}
\end{figure}

 In Sec.~(\ref{s2}), we have presented the hardware experimental analog circuit simulation results for the two-coupled 
Duffing oscillators. We have also performed an analog circuit simulation with $n=60$ using Pspice circuit simulator. We preferred 
the circuit simulator over hardware experiments due to the difficulty in the implementation of large size circuits on circuit boards. 
The various features of the multiple resonance and anti-resonance observed in the numerical simulation are also realized in the 
analog circuit simulation. For example, the emergence of multiple resonant peaks with increase of the number of coupled 
oscillators observed experimentally is shown in Fig. \ref{fig10}. The $n$ resonant peaks and $n-1$ anti-resonances are clearly 
visible for smaller values of $n$. 

\section{\label{s5}Conclusion}
In this paper, we have reported the occurrence of multiple resonance and anti-resonance in a system of $n$-coupled Duffing 
oscillators where the only the first oscillator is driven by a an external periodic force with a nearest neighbour coupling. 
In the case of uni-directionally coupled Duffing oscillators where also the first oscillator is the only one driven by the
external periodic force, the coupling is found to give rise to either an enhanced single resonance or just suppress the 
resonance depending upon the coupling strength \cite{r18}. One source for multi-resonance and anti-resonance is the type of 
coupling considered in the present work. Parametric anti-resonance \cite{r40,r20}, stochastic anti-resonance \cite{r35,r36} 
and coherence anti-resonance \cite{r23} have been found to occur in certain oscillators with a single degree of freedom. 
Investigation of resonance and anti-resonance in 
$n$-coupled version of such oscillators with the type of coupling analyzed in the present work may give
rise to new and interesting results. Another kind of systems where such a study has to be performed is in excitable 
systems such as FitzHugh-Nagumo equations \cite{r24}. As network models may represent many physical and biological
systems, it is also very important to identify the multiple resonance and anti-resonances in various network topologies.  
%
%
\begin{acknowledgements}
The work of RJ is supported by the University Grants Commission, Government of India in the form of 
Research Fellowship in Science for Meritorious Students. The work of KT forms a part of a Department 
of Science and Technology, Government of India sponsored project grant no. SR/S2/HEP-015/2010. MAFS acknowledges the 
financial support by the Spanish Ministry of Economy and Competitivity under project number FIS2013-40653-P.
\end{acknowledgements}


\end{document}